\documentclass[english]{article}
\usepackage[T1]{fontenc}
\usepackage[latin9]{inputenc}
\usepackage{amsbsy}
\usepackage{esint}

\makeatletter

\newcommand{\noun}[1]{\textsc{#1}}

\makeatother

\usepackage{babel}
\begin{document}

\title{The Relation Between Bayesian Fisher Information and Shannon Information
for Detecting a Change in a Parameter}

\author{Eric Clarkson}
\maketitle
\begin{abstract}
We derive a connection between performance of estimators the performance
of the ideal observer on related detection tasks. Specifically we
show how Shannon Information for the task of detecting a change in
a parameter is related to the Bayesian Fisher Information. We have
previously shown that this Shannon Information is related via an integral
transform to the Minimum Probability of Error on the same task. We
then outline circle of relations starting with this Minimum Probability
of Error and Ensemble Mean Squared Error via the Ziv-Zakai inequality,
then the Ensemble Mean Squared error for an estimator and the Bayesian
Fisher Information via the van Trees Inequality, and finally the Bayesian
Fisher Information and the Shannon Information for a detection task
via the work here.
\end{abstract}

\section{Introduction}

Fisher Information (FI) is normally thought of as related to the performance
of an estimator of a parameter in a statistical model for the given
data. The ordinary FI is related through the Cramer-Rao Bound (CRB)
to the variance of an unbiased estimator of the parameter. what we
will call the Bayesian FI (BFI) is related by the van Tress inequality
to the Ensemble Mean Squared Error (EMSE) for an estimator. On the
other hand we have shown previously that the FI is related to the
performance of the ideal observer on the task of detecting a small
change in the parameter, as measured by the area under the Receiver
Operating Characteristic (ROC) curve. Thus FI is related to performance
on a binary classification task. The Ziv-Zakai inequality relates
the EMSE of an estimator to the performance of the ideal observer
on the task of distinguishing two different values of the parameter
as measured by the minimum probability of error (MPE). Thus EMSE is
also related to performance on a binary classification task. In this
work we will derive further connections between performance of estimators
the performance of the ideal observer on related detection tasks.
Specifically we will show how Shannon Information (SI) for a certain
binary classification task is related to the BFI. We have previously
shown that this SI is related via an integral transform to the MPE
on the same task. We then have a circle of relations starting with
MPE and EMSE via Ziv-Zakai, then EMSE and BFI via van Trees, BFI and
SI via the work here, and finally SI and MPE via the aforementioned
integral transform. 

In Section 2 we have a brief review of FI and the CRB. Section 3 contains
a previously derived result relating the area ubder the ideal observer
ROC curve to FI when the task is to detect a small change in the parameter.
The reason for showing this result here is to emphasize the similarity
with the results below relating SI with BFI. In Section 4 we define
the BFI and briefly discuss the van Trees inequality. The Ziv-Zakai
inequality is presented in Section 5, and we have a short calculation
that derives a more compact form that is easily related to other results
here. Section 6 presents a derivation of the main result which relates
the conditional entropy for the task of detecting a small change in
a parameter to the BFI. In Section 7 we show the vector-parameter
version of the main result. The main result is reformulatd in terms
of SI in Section 8 and our conclusions are stated in Section 9.

\section{Fisher information}

We will consider a scalar parameter $\theta$ throughout. This parameter
takes values on the real line. There are also vector parameter versions
of everything. The data is a random vector $\mathbf{g}$generated
by a noisy imaging system, and has a conditional probability density
function (PDF) denoted by $pr\left(\mathbf{g}|\theta\right)$ . The
Fisher Information (FI) is then given by {[}1,2{]}

\begin{equation}
F\left(\theta\right)=\left\langle \left[\frac{d}{d\theta}\ln pr\left(\mathbf{g}|\theta\right)\right]^{2}\right\rangle _{\mathbf{g}|\theta}.
\end{equation}
Suppose that $\hat{\theta}\left(\mathbf{g}\right)$ is an estimator
that uses the data vector to produce an estimate of $\theta$. The
Cramer-Rao Bound (CRB) states that
\begin{equation}
var\left(\hat{\theta}\right)\geq\left[F\left(\theta\right)\right]^{-1}
\end{equation}
when
\begin{equation}
\left\langle \hat{\theta}\left(\mathbf{g}\right)\right\rangle _{\mathbf{g}|\theta}=\theta.
\end{equation}
In other words the variance of an unbiased estimator is bounded below
by the reciprocal of the FI. For the Bayesian FI there is the van
Trees inequality which is similar to the Cramer-Rao lower bound. This
will be discussed further below.

\section{FI and AUC}

We consider a classification task to classify $\mathbf{g}$ as a sample
drawn from from $pr\left(\mathbf{g}|\theta\right)$ or $pr\left(\mathbf{g}|\tilde{\theta}\right)$,
corresponding to hypotheses $H_{0}$ and $H_{1}$ respectively. The
ideal-observer AUC for this task computes the likelihood ratio {[}3{]}
\begin{equation}
\Lambda\left(\mathbf{g}\right)=\frac{pr\left(\mathbf{g}|\tilde{\theta}\right)}{pr\left(\mathbf{g}|\theta\right)}
\end{equation}
and compare to a threshold $\Lambda_{0}$ in order to make the decision.
If the likelihood ratio is greater than the threshold, then the decision
$D_{1}$ is made that $\mathbf{g}$ was drawn from from $pr\left(\mathbf{g}|\tilde{\theta}\right)$.
If the likelihood ratio is less than or equal to the threshold, then
the decision $D_{0}$ is made that $\mathbf{g}$ was drawn from from
$pr\left(\mathbf{g}|\theta\right)$. The True Positive Fraction (TPF)
is $Pr\left(D_{1}|H_{1}\right)$, the probability that the decision
$D_{1}$ is made when $H_{1}$ is true. The False Positive Fraction
(FPF) is $Pr\left(D_{1}|H_{0}\right)$, the probability that the decision
$D_{1}$ is made when $H_{0}$ is true. The Receiver Operating Characterisitc
(ROC) curve for this task plots TPF versus FPF as the threshold is
varied from $0$ to $\infty$. This is a concave curve that starts
at the point $\left(1,1\right)$ and ends at $\left(0,0\right)$.
The area under the ideal-observer ROC curve, the AUC, is a commonly
used figure of merit (FOM) for an imaging system on a classification
task. In this case we will denote this quantity by $AUC\left(\theta,\tilde{\theta}\right)$.
For the ideal observer the AUC is always between $0.5$ and $1$.
This area is related to the detectability $d\left(\theta,\tilde{\theta}\right)$
by the equation

\begin{equation}
AUC\left(\theta,\tilde{\theta}\right)=\frac{1}{2}+\frac{1}{2}\mathrm{erf}\left[\frac{1}{2}d\left(\theta,\tilde{\theta}\right)\right]
\end{equation}
The detectability is an equivalent FOM to the AUC and varies from
$0$ to $\infty$. In a previous publication we showed that that the
Taylor series expansion for the ideal-observer detectability in this
situation is given by {[}4,5,6{]}
\begin{equation}
d^{2}\left(\theta,\theta+\triangle\theta\right)=\left(\triangle\theta\right)^{2}F\left(\theta\right)+\ldots
\end{equation}
Our goal in this work is to derive a similar relation between the
Bayesian FI and the average Shannon Information (SI) for the classification
task that we have defined here.

\section{Bayesian FI and EMSE}

For the rest of this paper we will assume that we have a prior probability
$pr\left(\theta\right)$ on the parameter of interest. The Bayesian
FI is given by

\begin{equation}
F=\left\langle F\left(\theta\right)\right\rangle _{\theta}+\left\langle \left[\frac{d}{d\theta}\ln pr\left(\theta\right)\right]^{2}\right\rangle _{\theta}.
\end{equation}
For an alternate expression of the Bayesian FI we define the posterior
PDF $pr\left(\theta|\mathbf{g}\right)$ on $\theta$ by
\begin{equation}
pr\left(\theta|\mathbf{g}\right)=\frac{pr\left(\mathbf{g}|\theta\right)pr\left(\theta\right)}{pr\left(\mathbf{g}\right)},
\end{equation}
where
\begin{equation}
pr\left(\mathbf{g}\right)=\int_{-\infty}^{\infty}pr\left(\mathbf{g}|\theta\right)pr\left(\theta\right)d\theta.
\end{equation}
In terms of the posterior PDF we can show that
\begin{equation}
F=\left\langle \left\langle \left[\frac{d}{d\theta}\ln pr\left(\theta|\mathbf{g}\right)\right]^{2}\right\rangle _{\mathbf{g}|\theta}\right\rangle _{\theta}.
\end{equation}

The Ensemble Mean Squared Error (EMSE) for any estimator of $\theta$
is defined by 
\begin{equation}
EMSE=\left\langle \left\langle \left[\hat{\theta}\left(\mathbf{g}\right)-\theta\right]^{2}\right\rangle _{\mathbf{g}|\theta}\right\rangle _{\theta}
\end{equation}
The van Trees inequality states that {[}7,8{]}
\begin{equation}
EMSE\geq F^{-1}.
\end{equation}
In other words, the EMSE for any estimator of the parameter of interest
is bounded below by the reciprocal of the Bayesian FI. The van Tress
inequality is also called the Bayesian CRB. This is the reason for
calling $F$ the Bayesian FI. This inequality, much like the CRB,
relates the relevant version of the FI to performance on an estimation
task. On the other hand, the result described in Section 2 relates
the FI to a classification task, namely the detection of a small change
in the parameter of interest. We want to connect the Bayesian FI to
this same task, but before we do this we discuss a relation between
the EMSE and the detection task in Section 2 .

\section{Ziv-Zakai inequality (EMSE and MPE)}

When the prior $pr\left(\theta\right)$ is known, then we can define
prior probabilities for the two hypotheses $H_{0}$ and $H_{1}$ for
the detection task in Section 2 via
\begin{equation}
Pr_{0}\left(\theta,\tilde{\theta}\right)=\frac{pr\left(\theta\right)}{pr\left(\theta\right)+pr\left(\tilde{\theta}\right)}
\end{equation}
and 
\begin{equation}
Pr_{1}\left(\theta,\tilde{\theta}\right)=\frac{pr\left(\tilde{\theta}\right)}{pr\left(\theta\right)+pr\left(\tilde{\theta}\right)}.
\end{equation}
Then the minimum probabilty of error $P_{e}\left(\theta,\tilde{\theta}\right)$
on the this detection task is the probability of error for the ideal
observer when the threshold used for the likelihood ratio is given
by 
\begin{equation}
\Lambda_{0}=y\left(\theta,\tilde{\theta}\right)=\frac{Pr_{0}\left(\theta,\tilde{\theta}\right)}{Pr_{1}\left(\theta,\tilde{\theta}\right)}=\frac{pr\left(\theta\right)}{pr\left(\tilde{\theta}\right)}.
\end{equation}
The minimum probability of error is $P_{e}\left(\theta,\tilde{\theta}\right)=\left(FPF\right)Pr_{0}+\left(1-TPF\right)Pr_{1}$
since $1-TPF$ is the False Negative Fraction (FNF) $Pr\left(D_{0}|H_{1}\right)$.
Of course, in this expression, $TPF$ and $FPF$ also depend on the
pair $\left(\theta,\tilde{\theta}\right)$. The Ziv-Zakai inequality,
in its standard form, states that {[}9,10{]}

\begin{equation}
EMSE\geq\frac{1}{2}\int_{0}^{\infty}\int_{-\infty}^{\infty}\left[pr\left(\theta\right)+pr\left(\theta+x\right)\right]P_{e}\left(\theta,\theta+x\right)d\theta xdx
\end{equation}

We will show that this inequality can be written in an alternate form
which connects it to the classification task in Section 2. We start
with the substitution $\tilde{\theta}=\theta+x$ so that we have
\begin{equation}
EMSE\geq\frac{1}{2}\int_{-\infty}^{\infty}\int_{\theta}^{\infty}\left[pr\left(\theta\right)+pr\left(\tilde{\theta}\right)\right]P_{e}\left(\theta,\tilde{\theta}\right)\left(\tilde{\theta}-\theta\right)d\tilde{\theta}d\theta.
\end{equation}
Due to the range of integration for $\tilde{\theta}$ we may also
write this as
\begin{equation}
EMSE\geq\frac{1}{2}\int_{-\infty}^{\infty}\int_{\theta}^{\infty}\left[pr\left(\theta\right)+pr\left(\tilde{\theta}\right)\right]P_{e}\left(\theta,\tilde{\theta}\right)\left|\tilde{\theta}-\theta\right|d\tilde{\theta}d\theta
\end{equation}
Interchanging the order of integration we have
\begin{equation}
EMSE\geq\frac{1}{2}\int_{-\infty}^{\infty}\int_{-\infty}^{\tilde{\theta}}\left[pr\left(\theta\right)+pr\left(\tilde{\theta}\right)\right]P_{e}\left(\theta,\tilde{\theta}\right)\left|\tilde{\theta}-\theta\right|d\theta d\tilde{\theta}
\end{equation}
Now we note that $P_{e}\left(\theta,\tilde{\theta}\right)=P_{e}\left(\tilde{\theta},\theta\right)$
since the two tasks described by the pairs $\left(\theta,\tilde{\theta}\right)$
and $\left(\tilde{\theta},\theta\right)$ are the same task. Therefore
we may interchange the variables $\theta$ and $\tilde{\theta}$ and
use the symmetry of all of the functions in the integrand to arrive
at
\begin{equation}
EMSE\geq\frac{1}{2}\int_{-\infty}^{\infty}\int_{-\infty}^{\theta}\left[pr\left(\theta\right)+pr\left(\tilde{\theta}\right)\right]P_{e}\left(\theta,\tilde{\theta}\right)\left|\tilde{\theta}-\theta\right|d\tilde{\theta}d\theta.
\end{equation}
Now combining the second and fourth inequalities in this chain we
have
\begin{equation}
EMSE\geq\frac{1}{4}\int_{-\infty}^{\infty}\int_{-\infty}^{\infty}\left[pr\left(\theta\right)+pr\left(\tilde{\theta}\right)\right]P_{e}\left(\theta,\tilde{\theta}\right)\left|\tilde{\theta}-\theta\right|d\tilde{\theta}d\theta
\end{equation}
We separate the right hand side into two integrals and again use the
symmetry in $\theta$ and $\tilde{\theta}$ to find out that the two
integrals are the same. Therefore we have the final result
\begin{equation}
EMSE\geq\frac{1}{2}\int_{-\infty}^{\infty}\int_{-\infty}^{\infty}pr\left(\theta\right)P_{e}\left(\theta,\tilde{\theta}\right)\left|\tilde{\theta}-\theta\right|d\tilde{\theta}d\theta.
\end{equation}
Finally, the integral on the right can be written as an expectation:
\begin{equation}
EMSE\geq\frac{1}{2}\left\langle \int_{-\infty}^{\infty}P_{e}\left(\theta,\tilde{\theta}\right)\left|\tilde{\theta}-\theta\right|d\tilde{\theta}\right\rangle _{\theta}.
\end{equation}
This formulation of the Ziv-Zakai inequality makes it clear that the
worst case scenario, from the EMSE persepctive, is when the minimum
proabablity of error $P_{e}\left(\theta,\tilde{\theta}\right)$ on
the classification task in Section 2 is significant for relatively
large values of $\left|\tilde{\theta}-\theta\right|$. In the next
section we will complete the circle by relating the Bayesian FI to
performance on the classification task in Section 2 as measured by
the average SI.

\section{SI and the Bsyesian FIM}

For the purposes of this section we will use the notation

\begin{equation}
\Lambda\left(\mathbf{g}|\theta,\tilde{\theta}\right)=\frac{pr\left(\mathbf{g}|\tilde{\theta}\right)}{pr\left(\mathbf{g}|\theta\right)}
\end{equation}
for the likelihood ratio for the classification task in Section 2.
Keeping in mind that $\Lambda=\Lambda\left(\mathbf{g}|\theta,\tilde{\theta}\right)$
and $y=y\left(\theta,\tilde{\theta}\right)$, we can write the SI
for this classification task as {[}11{]}
\begin{equation}
I\left(y\right)=\frac{1}{1+y}\left\langle \ln\left[\frac{\Lambda\left(1+y\right)}{\Lambda+y}\right]\right\rangle _{\mathbf{g}|\tilde{\theta}}+\frac{y}{1+y}\left\langle \ln\left[\frac{1+y}{\Lambda+y}\right]\right\rangle _{\mathbf{g}|\theta}.
\end{equation}
If we introduce a binary random variable $X$ such that $X=1$ with
probability $Pr_{1}\left(\theta,\tilde{\theta}\right)$, and $X=0$
with probability $Pr_{0}\left(\theta,\tilde{\theta}\right)$, then
the entropy of $X$ is given by
\begin{equation}
H\left(y\right)=-\frac{1}{1+y}\ln\left(\frac{1}{1+y}\right)-\frac{y}{1+y}\ln\left(\frac{y}{1+y}\right).
\end{equation}
The conditional entropy for $X$ given $\mathbf{g}$ is then defined
by $C_{e}\left(y\right)=H\left(y\right)-I\left(y\right).$ The non-standard
notation for conditional entropy here was introduced in an earlier
publication that related conditional entropy to MPE via an integral
transform in the variable $y$. After some algebra we have the expression
\[
C_{e}\left(y\right)=\frac{1}{1+y}\left\langle \ln\left(\frac{\Lambda+y}{\Lambda}\right)\right\rangle _{\mathbf{g}|\tilde{\theta}}+\frac{y}{1+y}\left\langle \ln\left(\frac{\Lambda+y}{y}\right)\right\rangle _{\mathbf{g}|\theta}.
\]
We are interested in the limiting value of the conditional entropy
as $\tilde{\theta}$ approaches $\theta$. After some more simplification
involving the definition of the likelihood ratio we can write the
conditioal entropy for the task in Section 2 in the form

\[
C_{e}\left(\theta,\tilde{\theta}\right)=C_{e}\left(y\right)=\frac{1}{1+y}\left[\left\langle \left(\Lambda+y\right)\ln\left(\Lambda+y\right)\right\rangle _{\mathbf{g}|\theta}-\left\langle \Lambda\ln\Lambda\right\rangle _{\mathbf{g}|\theta}-y\ln y\right].
\]
Note that only expectations under hypothesis $H_{0}$ are needed to
compute the conditional entropy, and hence the SI. 

In order to keep the notational complications to a minimum we introduce
the function
\[
C_{e}'\left(\theta\right)=\left.\frac{d}{d\tilde{\theta}}C_{e}\left(\theta,\tilde{\theta}\right)\right|_{\tilde{\theta}=\theta}.
\]
We also will use the following notation

\begin{equation}
\Lambda'=\Lambda'\left(\mathbf{g}|\theta\right)=\left.\frac{d}{d\tilde{\theta}}\Lambda\left(\mathbf{g}|\theta,\tilde{\theta}\right)\right|_{\tilde{\theta}=\theta}=\frac{pr'\left(\mathbf{g}|\theta\right)}{pr\left(\mathbf{g}|\theta\right)},
\end{equation}
where the prime on the far right indicates a derivative with respect
to $\theta$. Similarly we will write
\begin{equation}
y'=y'\left(\theta\right)=\left.\frac{d}{d\tilde{\theta}}y\left(\theta,\tilde{\theta}\right)\right|_{\tilde{\theta}=\theta}=\frac{pr'\left(\theta\right)}{pr\left(\theta\right)}.
\end{equation}
Now, using the fact that, when $\tilde{\theta}=\theta$, we have $y=1$
and $\Lambda=1$, 
\[
C_{e}'\left(\theta\right)=-\frac{y'}{4}2\ln2+\frac{1}{2}\left[\left\langle \left(\Lambda'+y'\right)\ln2+\left(\Lambda'+y'\right)-\Lambda'\right\rangle _{\mathbf{g}|\theta}-y'\right].
\]
Now we use the fact that $\left\langle \Lambda'\right\rangle _{\mathbf{g}|\theta}=0$
to give us

\[
C_{e}'\left(\theta\right)=-\frac{y'}{4}2\ln2+\frac{1}{2}\left[y'\ln2\right]=0.
\]
This result is to be expected since, as a function of $\tilde{\theta}$,
the conditional entropy $C_{e}\left(\theta,\tilde{\theta}\right)$
is maximized when $\tilde{\theta}=\theta$. 

Now we will look at second derivative terms. We define
\[
C_{e}''\left(\theta\right)=\left.\frac{d^{2}}{d\tilde{\theta^{2}}}C_{e}\left(\theta,\tilde{\theta}\right)\right|_{\tilde{\theta}=\theta}.
\]
We also will use the following notation

\begin{equation}
\Lambda''=\Lambda''\left(\mathbf{g}|\theta\right)=\left.\frac{d^{2}}{d\tilde{\theta^{2}}}\Lambda\left(\mathbf{g}|\theta,\tilde{\theta}\right)\right|_{\tilde{\theta}=\theta}=\frac{pr''\left(\mathbf{g}|\theta\right)}{pr\left(\mathbf{g}|\theta\right)},
\end{equation}
and
\begin{equation}
y''=y''\left(\theta\right)=\left.\frac{d^{2}}{d\tilde{\theta}^{2}}y\left(\theta,\tilde{\theta}\right)\right|_{\tilde{\theta}=\theta}=\frac{pr''\left(\theta\right)}{pr\left(\theta\right)}.
\end{equation}
In order to ease the mental computations for the reader we note that,
for any variable $z$ that depends on $\theta$, we have 
\[
\frac{d}{d\theta}z\ln z=z'\left(1+\ln z\right)
\]
and
\[
\frac{d^{2}}{d\theta^{2}}z\ln z=z''\left(1+\ln z\right)+\frac{\left(z'\right)^{2}}{z}.
\]
Also note that we use
\[
\frac{d}{d\theta}\frac{1}{1+y}=-\frac{y'}{\left(1+y\right)^{2}}
\]
and
\[
\frac{d^{2}}{d\theta^{2}}\frac{1}{1+y}=\frac{2\left(y'\right)^{2}}{\left(1+y\right)^{3}}-\frac{y''}{\left(1+y\right)^{2}}.
\]
Now using the Liebniz rule for second derivatives we have the expansion
$C_{e}''\left(\theta\right)=A+2B+C,$where the terms on the right
are given by
\[
A=\left\langle \left[\frac{\left(y'\right)^{2}-y''}{4}\right]\left(2\ln2\right)\right\rangle _{g|\theta},
\]
\[
2B=\left\langle -\frac{y'}{2}\left[\left(1+\ln2\right)\left(\Lambda'+y'\right)-\Lambda'-y'\right]\right\rangle _{g|\theta}
\]
and
\[
C=\left\langle \frac{1}{2}\left[\left(\Lambda''+y''\right)\left(1+\ln2\right)+\frac{1}{2}\left(\Lambda'+y'\right)^{2}-\Lambda''-\left(\Lambda'\right)^{2}-y''-\left(y'\right)^{2}\right]\right\rangle _{g|\theta}.
\]
These three terms simplify algebraically to
\[
A=\left\langle \left[\left(y'\right)^{2}-y''\right]\frac{\ln2}{2}\right\rangle _{g|\theta},
\]
\[
2B=\left\langle -\frac{\ln2}{2}\left[\Lambda'y'+\left(y'\right)^{2}\right]\right\rangle _{g|\theta}
\]
and
\[
C=\left\langle \left(\Lambda''+y''\right)\frac{\ln2}{2}+\frac{1}{2}\Lambda'y'-\frac{1}{4}\left(\Lambda'\right)^{2}-\frac{1}{4}\left(y'\right)^{2}\right\rangle _{g|\theta}.
\]
Now we use the fact that $\left\langle \Lambda^{''}\right\rangle _{g|\theta}=0$
and $\left\langle \Lambda'y'\right\rangle _{g|\theta}=y'\left\langle \Lambda'\right\rangle _{g|\theta}=0$
to arrive at
\[
C_{e}''\left(\theta\right)=\left\langle -\frac{1}{4}\left(\Lambda'\right)^{2}-\frac{1}{4}\left(y'\right)^{2}\right\rangle _{g|\theta}=-\frac{1}{4}\left[\left\langle \left(\Lambda'\right)^{2}\right\rangle _{g|\theta}+\left(y'\right)^{2}\right].
\]
If we use the prior to average over $\theta$ we then have
\[
\left\langle C_{e}''\left(\theta\right)\right\rangle _{\theta}=-\frac{1}{4}\left\langle \left\langle \left(\Lambda'\right)^{2}\right\rangle _{g|\theta}+\left(y'\right)^{2}\right\rangle _{\theta}=-\frac{1}{4}F.
\]
Since $\left\langle C_{e}\left(\theta,\theta\right)\right\rangle _{\theta}=\left\langle \ln2\right\rangle _{\theta}=\ln2$
we have the Taylor expansion of the average conditional entropy for
the task in Section 2:
\[
\left\langle C_{e}\left(\theta,\theta+\triangle\theta\right)\right\rangle _{\theta}=\ln2-\frac{\left(\triangle\theta\right)^{2}}{4}F+\ldots
\]
Thus the Bayesian FI gives the lowest order approximation for the
average conditional entropy when the task is detecting a small change
in the parameter of interest. 

The Ziv-Zakai inequality provides a relation between the MPE for the
task of using the data vector $\mathbf{g}$ to detect a change in
a parameter $\theta$ in the conditional PDF $pr\left(\mathbf{g}|\theta\right)$,
and the EMSE for the task of using $\mathbf{g}$ to estimate a parameter
$\theta$ in the conditional PDF $pr\left(\mathbf{g}|\theta\right)$.
The van Trees inequality relates this EMSE the Bayesian FI. We have
now found a relationship between the Bayesian FI and the average conditional
entropy for the detection task. This completes the circle since there
is an integral relation between the conditionaal entropy for any classification
task and the MPE for the same task {[}12,13{]}. This reinforces the
idea that there is a strong mathematical relationship between the
performance of an imaging system on estimation tasks and the performance
of the same system on the detection of a small change in the parameter
of interest.

\section{Vector version}

For athe vector version of this last result we assume a vector parameter
$\boldsymbol{\theta}$ and a perturbation in this parameter of the
form $t\mathbf{u}$, where $\mathbf{u}$ is a unit vector in parameter
space. We then have a Taylor series expansion in the varaible $t$
given by

\[
\left\langle C_{e}\left(\boldsymbol{\theta},\boldsymbol{\theta}+t\boldsymbol{u}\right)\right\rangle _{\boldsymbol{\theta}}=\ln2-\frac{t^{2}}{4}\mathbf{u}^{\dagger}\mathbf{F}\mathbf{u}+\ldots
\]
in this equation the Bayesian Fisher Information Matrix (FIM) is given
by
\begin{equation}
\mathbf{F}=\left\langle \mathbf{F}\left(\boldsymbol{\theta}\right)\right\rangle _{\theta}+\left\langle \left[\nabla_{\boldsymbol{\theta}}\ln pr\left(\boldsymbol{\theta}\right)\right]\left[\nabla_{\boldsymbol{\theta}}\ln pr\left(\boldsymbol{\theta}\right)\right]^{\dagger}\right\rangle _{\boldsymbol{\theta}},
\end{equation}
where the ordinary FIM is defined by 
\begin{equation}
\mathbf{F}\left(\boldsymbol{\theta}\right)=\left\langle \left[\nabla_{\boldsymbol{\theta}}\ln pr\left(\mathbf{g}|\boldsymbol{\theta}\right)\right]\left[\nabla_{\boldsymbol{\theta}}\ln pr\left(\mathbf{g}|\boldsymbol{\theta}\right)\right]^{\dagger}\right\rangle _{\mathbf{g}|\boldsymbol{\theta}}.
\end{equation}
There are vector versions of the van Trees inequality and the Ziv-Zakai
inequality so the circle of relations between EMSE, MPE and Bayesian
FIM also exists for vector parameters..

\section{Taylor series expansion for SI}

To get a Taylor series expansion for SI similar to the one we have
derived for conditional entropy we must find the expansion for $H\left(y\right)$.
This function can be written in the form
\[
H\left(y\right)=\frac{1}{1+y}\left[\left(1+y\right)\ln\left(1+y\right)-y\ln\left(y\right)\right].
\]
Using notation similar to that in Section 5 we have
\[
H'\left(\theta\right)=-\frac{y'}{4}\left(2\ln2\right)+\frac{1}{2}\left(y'\ln2+y'-y'\right)=0.
\]
For the second derivative we can write $H''\left(\theta\right)=a+2b+c$,
with
\[
a=\left[\frac{\left(y'\right)^{2}}{4}-\frac{y''}{4}\right]\left(2\ln2\right),
\]
\[
2b=-2\left(\frac{y'}{4}\right)\left(y'\ln2+y'-y'\right)
\]
and
\[
c=\left(\frac{1}{2}\right)\left\{ y''\left[1+\ln2\right]+\frac{\left(y'\right)^{2}}{2}-y''-\left(y'\right)^{2}\right\} .
\]
Combining terms we have
\[
H''\left(\theta\right)=-\frac{\left(y'\right)^{2}}{4}.
\]
Now we may combine these computations with the expansion in Section
5 and find that
\[
I''\left(\theta\right)=\frac{1}{4}\left\langle \left(\Lambda'\right)^{2}\right\rangle _{g|\theta}=\frac{1}{4}F\left(\theta\right)
\]
This gives us another relation beween the FI and the detection task
in Section 2. The Taylor series expansion for $I\left(\theta,\theta+\triangle\theta\right)$
now has the form
\[
I\left(\theta,\theta+\triangle\theta\right)=\frac{\left(\triangle\theta\right)^{2}}{4}F\left(\theta\right)+\ldots
\]
Thus the SI and the ideal-observer detectability are essentially the
same FOM when we are trying to detect a small change in a parameter. 

After using the prior on $\theta$ to compute an expectation we have
\[
\left\langle I\left(\theta,\theta+\triangle\theta\right)\right\rangle _{\theta}=\frac{\left(\triangle\theta\right)^{2}}{4}\left\langle F\left(\theta\right)\right\rangle _{\theta}+\ldots
\]
Thus the quantity $\left\langle F\left(\theta\right)\right\rangle _{\theta}$
also has an interpretation in terms of the average SI for the detection
task in Section 2. 

\section{Conclusion}

Fisher Information (FI) is almost always thought of in terms of the
cramer-Rao Bound. What we call the Bayesian Fisher Information is
usually thought of in terms of the van Trees inequality. We have shown
previously that Fisher Information is related to the performance of
the ideal observer on the task of detecting a small change in the
parameter, as measured by the area under the Receiver Operating Characteristic
curve. Thus Fisher Information is related to the ideal performance
on a binary classification task. The Ziv-Zakai inequality relates
the Ensemble Mean Squared Error of an estimator to the performance
of the ideal observer on the task of distinguishing two different
values of the parameter, as measured by the Minimum Probability of
Error. In this work we showed how Shannon Information for a certain
binary classification task is related to the Bayesian Fisher Information.
Since we have previously shown that this Shannon Information is related
via an integral transform to the Minimum Probability of Error on the
same task, we then have a circle of relations relating Minimum Probability
of Error on a detection tsak, Ensemble Mean Squared Error of an Estimator,
Bayesian Fisher Information, and Shannon Information for the detection
task.

One of the useful results of this work is that Tsak-Based Shannon
Information (TSI) for the detection of a small change in a parameter
and the ideal-observer AUC for the same task at essentially equivalent
figures of merit for a given imaging system. Optimizing asystem for
one of these quantities will also optimoze it for the other. Another
point of interest is that, when the parameter is a vector, the interpretation
of the Bayesian Fisher Information Matrix in terms of Shannon Information
does not require inversion of the matrix. This is similar to the connection
betwee the Fisher Information Matrix and the ideal-observer AUC derived
previously. Finally, both of these relations are the first terms iof
a Taylor series expansion, and not inequalities as the Cramer-Rap
Bound and the van Trees inequality are. They thus have the potential
to provide good approximations to the relevant TSI or AUC values.

\end{document}